\documentclass[preprint,prd,aps,nofootinbib]{revtex4}
\usepackage{graphicx}

\begin{document}
\title
{\large Soft Nonfactorizable Contribution to $\bar{B}^0\to
D^{0}\pi^{0}$}

\author{Jian-Ying Cui}
\email[Email: ]{cjy@ytu.edu.cn}
\address{Department of Physics,
Yantai University,\\ Yantai, 264005, China}

\author{Zuo-Hong Li}
\email[Email: ]{lizh@ytu.edu.cn}
\address{Department of Physics,
Yantai University, Yantai, 264005, China \\ and \\ CCAST (World
Laboratory), P.O.Box 8730,\\ Beijing 100080, China}
\date{\today}
\begin{abstract}

For the color-suppressed $\bar{B}^0\longrightarrow D^{0}\pi^{0}$
decay, nonfactorizable contributions are expected to be leading
and naive factorization description breaks down. We study $1/m_b$
power-suppressed nonfactorizable effect in
$\bar{B}^0\longrightarrow D^{0}\pi^{0}$, which is due to soft
exchange between the emitted heavy-light quark pair and $B\pi$
system, in the framework of QCD light-cone sum rules (LCSR). The
resulting correction to the decay amplitude is found to be
numerically comparable with the corresponding factorizable piece,
estimated at about $(50-110)\%$ of the latter. The relevant
parameter $a_2$ receives a positive number contribution, due to
the factorizable correction and the power-suppressed soft effect.
Our findings would be crucial to phenomenological understanding of
the $\bar{B}^0\longrightarrow D^{0}\pi^{0}$ decay.
\end{abstract}

\maketitle

\section{introduction}

``Naive'' factorization \cite{nf}, or ``generalized''
factorization \cite{gf} developed subsequently, had been viewed as
a simple but predictive model for two-body hadronic decay of heavy
mesons prior to presentation of QCD factorization \cite{qcdfa}. At
present, it is known to us that for a large class but not all of
two-body nonleptionic B decays, QCD factorization can furnish a
rigorous theoretical basis for the naive factorization assumption
of the hadronic matrix elements. Some of examples, for which the
naive factorization holds up to power corrections in
$\Lambda_{QCD}/m_b$ and $\alpha_s$, are the charmless decays $B\to
\pi \pi,\pi K$ and the class-1 charmed decays $\bar{B}^0\to
D^{(*)+}\pi^-$ (relevant to the parameter $a_1$). In the heavy
quark limit $m_b\longrightarrow \infty$, but to all orders of
perturbation theory, this type of processes can be systematically
computed in terms of convolutions of hard-scattering kernels with
the leading twist light-cone distribution amplitudes of the
corresponding emitted light mesons. Such a treatment is based on
color-transparency argument \cite{bjorken} that for the limit in
question the momentum carried by an light meson which is directly
emitted off the relevant weak vertex is so large that it has not
sufficient time to exchange soft gluons with the system including
the decaying B meson and the produced meson picking up a spectator
quark. Typical examples for which QCD factorization does not apply
are the class-2 charmed decays $\bar{B}^0\to D^{(*)0}\pi^0$
(usually called color-suppressed decays for the reason that the
relevant phenomenological parameters $a_2(D\pi)$ is of ${\cal
O}(1/N_c)$ in the large $N_c$ accounting \cite{Buras}). The reason
is that unlike the aforementioned case where a light meson is
emitted, the heavy $D^{(*)0}$ meson, as an emitted particle which
is neither small ( $\sim 1/\Lambda_{QCD}$) nor fast, is produced
in the color-suppressed decays and so can not be decoupled from
the $B\pi$ system. This indicates clearly that the factorization
contributions to $\bar{B}^0\to D^{(*)0}\pi^0$ don't provide a
leading result, for all that they are present, and nonfactorizable
soft contributions, for example, the charge-exchanging
rescattering processes \cite{Donoghue} from the dominate class-1
channel, dominate in such decays. Hence, at the present a
theoretical understanding is not accessible for the
color-suppressed decays $\bar{B}^0\to D^{(*)0}\pi^0$ and also for
the class-3 decays, say, $B^-\to D^{(*)0}\pi^-$ to a certain
extend, although in the latter case the class-2 amplitudes are
predicted to be power-suppressed with respective to the
corresponding class-1 ones in QCD factorization. It is quite a
challenge to give a consistent theoretical explanation of the data
on $B\to D^{(*)}\pi$, after new experimental observations $ {\cal
B}(\bar{B}^0 \rightarrow D^0\pi^0) = (2.74^{+0.36}_{-0.32} \pm
0.55) \cdot 10^{-4}$ \cite{CLEO} and $ {\cal B}(\bar{B}^0
\rightarrow D^0\pi^0) = (3.1 \pm 0.4 \pm 0.5) \cdot 10^{-4}$
\cite{Belle} are announced respectively by the CLEO and Belle
Collaborations. While there exist the attempts to quantitatively
understand the color-suppressed decays $\bar{B}^0\to
D^{(*)0}\pi^0$ in perturbative QCD(pQCD) \cite{Keum} and
soft-collinear effective theory (SCET) \cite{scet}, the
model-independent discussions \cite{neubert, xing, cheng} can help
to get interesting suggestions about the magnitudes and relative
phase of the parameters $a_{1,2}$. A couple of important findings
can be summarized as follows: (1) A sizable relative strong
interaction phase is expected between class-1 and class-2 $B\to
D^{*}\pi$ decay amplitudes \cite{neubert}. (2) The parameter
$|a_2|$ is extracted to be $|a_2(D\pi)|\sim 0.35-0.60$ and
$|a_2(D^*\pi)|\sim 0.25-0.50$ from the data \cite{cheng}. (3)
Several types of possible power corrections to $a_1$ parameter
have been estimated and found to be small, a near-universal value
$|a_1|\approx 1.1$ observed experimentally is now put on a firm
footing \cite{neubert}.

Now we are not able to give a reliable theoretical interpretation
for the first two observations, because of the unknown leading
nonperturbative effects involved in the parameter $a_2(D\pi)$.
However, the $1/m_b$ power-suppressed nonfactorizable
contributions to $a_2(D\pi)$, which come from soft exchange
between the emitted heavy-light quark pair and the $B\pi$ system,
would be expected to be much more important than in the case of
$B\to \pi \pi, \pi K$ and thus a reliable estimate of such effect
is crucial.

Earlier discussion on the power-suppressed contribution to
$\bar{B}^0\to D^0\pi^0$ in the QCD light-cone sum rule(LCSR)
approach \cite{LCSR} can be found in Ref. \cite{Halperin}. Also,
in the framework of a generalized QCD LCSR \cite{Kh} the similar
effects have been estimated for some of the other important $B$
decays \cite{Kh, LCSR1, LCSR2}. In this paper, we intend to apply
the generalized QCD LCSR approach to estimate the soft effect in
the color-suppressed $\bar{B}^0\to D^0\pi^0$ decay and then
compare the result yielded with the naive factorization
contribution.

This paper is organized as follows: the following Section contains
a detailed derivation of LCSR's for the power-suppressed soft
contribution to the $\bar{B}^0\longrightarrow D^{0}\pi^{0}$ decay
amplitude and the numerical results. The last Section is devoted
to discussion and conclusion.

\section{LSCR's for soft nonfactorizable effect}

The relevant effective weak Hamiltonian for the
$\bar{B}^0\longrightarrow D^{0}\pi^{0}$ decay is written as
\cite{Hamiltonian}
\begin{equation}
{\cal H}_{\rm W} = {G_F\over\sqrt{2}}\, V_{cb}V_{ud}^*
\Big[C_1(\mu)O_1(\mu)+C_2(\mu)O_2(\mu)\Big],
\label{eq:hamilton1}
\end{equation}
where $C_{1,2}$ are the Wilson coefficients, $V_{ij}$ the CKM
matrix elements and $O_{1,2}$ the four-quark operators given by
\begin{eqnarray}
O_1= (\bar c\Gamma^\mu u)(\bar d\Gamma_\mu b),~~~~~ O_2= (\bar
d\Gamma^\mu u)(\bar c\Gamma_\mu b),
\end{eqnarray}
with $\Gamma_\mu=\gamma_\mu(1- \gamma_5)$. Further, the use of
Fierz transformation can make (\ref{eq:hamilton1}) rewritten as
\begin{eqnarray}
{\cal H}_{\rm W} = {G_F\over\sqrt{2}}\, V_{cb}V_{ud}^*
\Big[(C_1(\mu)+\frac{1}{3}C_2(\mu))O_1(\mu)+2
C_2(\mu)\tilde{O}_1(\mu)\Big]\;,
 \label{eq:hamilton}
\end{eqnarray}
where
\begin{eqnarray}
\tilde{O}_1= (\bar c\frac{\lambda_a}{2}\Gamma_\mu u)(\bar
d\frac{\lambda_a}{2}\Gamma^\mu b)
\end{eqnarray}
with $\lambda_a$ being the color $SU(3)$ matrices.

Among the non-leading part of the decay amplitude $\langle D^0
\pi^0|{\cal H}_{\rm W}|\bar{B}^0\rangle$ is the factorizable and
power-suppressed soft contribution. We can parameterize it as
\begin{eqnarray}
{\cal A}_{NL}(\bar{B}^0\longrightarrow D^0\pi^0) &=&{\cal
A}_F(\bar{B}^0\longrightarrow D^0\pi^0)+{\cal
A}_S(\bar{B}^0\longrightarrow D^0\pi^0)\\ \nonumber
&=& -i
\frac{G_F}{2}V_{cb}V^*_{ud}m^2_B f_D F_0^{B\pi}(m^2_D)a_2^{NL}.
\end{eqnarray}
where $F_0^{B\pi}$ is the $B\to\pi$ form factor, $f_D$ the $D$
meson decay constant, ${\cal A}_F(\bar{B}^0\longrightarrow
D^0\pi^0)$ and ${\cal A}_S(\bar{B}^0\longrightarrow D^0\pi^0)$
express the factorizable and power-suppressed soft contributions
respectively:
\begin{equation}
{\cal A}_F(\bar{B}^0\longrightarrow D^0\pi^0)=-i
\frac{G_F}{2}V_{cb}V^*_{ud}m^2_B f_D
F_0^{B\pi}(m^2_D)(C_1+\frac{C_2}{3}),
\label{eq:factorizable}
\end{equation}
\begin{equation}
{\cal A}_S(\bar{B}^0\longrightarrow
D^0\pi^0)=\frac{G_F}{\sqrt{2}}V_{cb}V^*_{ud}(2C_2)\langle D^0
\pi^0|\tilde{O_1}|\bar{B}^0\rangle_S,
\label{eq:softnonfctorizable}
\end{equation}
and the parameter $a_2^{NL}$ is defined as
\begin{equation}
a_2^{NL}=(C_1+\frac{C_2}{3})\left(1+\frac{{\cal
A}_S(\bar{B}^0\longrightarrow D^0\pi^0)}{{\cal A
}_F(\bar{B}^0\longrightarrow D^0\pi^0)}\right),
\end{equation}

For a quantitative estimate of the nonfactorizable matrix element
$\langle D^0 \pi^0|\tilde{O_1}|\bar{B}^0\rangle$, we make use of
the generalized QCD LCSR method developed in \cite{Kh}. We start
with the correlation function:
\begin{equation}
F_{\alpha}(p,q,k) = i^2 \, \int d^4 x e^{-i(p+q)x} \int d^4 y
e^{i(p-k)y} \langle \pi^0(q) | T \{ j_{\alpha}^{D}(y)
{\tilde{O}_1}(0) j_5^{B}(x) \} | 0 \rangle \, , \label{eq:cor1}
\end{equation}
where $j_{\alpha}^{D} = \bar{u} \gamma_{\alpha}\gamma_5 c$ and
$j_5^{B} = m_b \bar{b} i \gamma_5 d$ are currents interpolating
the $D^0$ and $\bar{B}^{0}$ meson fields respectively. The
correlator is a function of three independent momenta chosen to be
$q$, $p-k$ and $k$. An important point of this method is to
introduce a fictitious unphysical momentum $k$. Consequently, in
the correlator the quark states before and after the $b$-quark
decay will have different four-momentum, and thus one avoids a
continuum of light parasitic contributions in the $B$ channel. Of
course, the unphysical quantity must disappear from the
$\bar{B}^0\longrightarrow D^{0}\pi^{0}$ ground state contribution
in the dispersion integral. This can be guaranteed, as will be
seen, by picking out a reasonable kinematical region for which the
LCSR calculation is effective.

The kinematical decomposition of the correlation function
(\ref{eq:cor1}) can proceed in the following form:
\begin{equation}
F_\alpha= (p-k)_\alpha F^{(p-k)} + q_\alpha F^{(q)} + k_\alpha
F^{(k)} + \epsilon_{\alpha\beta\lambda\rho}q^\beta p^\lambda
k^\rho F^{(\varepsilon)}\,,
 \label{eq:cor2}
\end{equation}
Here, $F^{i}$'s are scalar functions of 6 independent Lorentz
invariants, which are chosen to be: $P^2=(p+q-k)^2\, ,p^2\, ,
q^2\, , (p+q)^2\, ,k^2$ and $(p-k)^2$. Using the Operator Product
Expansion (OPE) near the light-cone $x^2\sim y^2\sim (x-y)^2\sim
0$, the correlator (\ref{eq:cor1}) is calculable. For the
calculation to go effectively, the momenta squared $P^2,(p+q)^2$
and $(p-k)^2$ have to be taken spacelike and large in order to
stay far away from the hadronic thresholds in both the $B$ and $D$
channels. Furthermore, a simple and possible choice, for the
external momentum squared $k^2$ and kinematical invariant $p^2$,
is to let $k^2=0$ and $p^2=m_D^2$, $m_D$ being the mass of $D^0$.
The pion is taken on shell and we set $q^2=0$. Altogether, the
kinematical region used for our LCSR calculation is
\begin{equation}
q^2 = k^2 = 0, \; p^2 = m_{D}^2, \; |(p-k)|^2 \gg \Lambda_{QCD},
|(p+q)|^2 \gg\Lambda_{QCD},  |P|^2 \gg \Lambda_{QCD} \,.
\label{eq:region}
\end{equation}

In this region the light-cone OPE is applicable to the correlator
(\ref{eq:cor1}) and the result can be expressed in the form of
hard scattering amplitudes convoluted with the pion light-cone
distribution amplitudes. We note that in (\ref{eq:cor2}) the
relevant invariant amplitude to our desire is only $F^{(p-k)}$.
The QCD result for $F^{(p-k)}$ is, in a general form of dispersion
relation, expressed as
\begin{equation}
F_{QCD}^{(p-k)}((p-k)^2,(p+q)^2,P^2) = \frac{1}{\pi}
\int_{m_c^2}^{\infty} ds \frac{{\rm Im}_s F^{(p-k)}_{QCD}
(s,(p+q)^2,P^2,p^2)}{s - (p-k)^2} \, . \label{eq:qcd}
\end{equation}

On the other hand, we can obtain a corresponding dispersion
relation in the hadronic level. By inserting in the right hand
side of (\ref{eq:cor1}) a complete set of hadronic states with
$D^0$ quantum numbers, we get:
\begin{equation}
F^{(p-k)}((p-k)^2,(p+q)^2,P^2,p^2)= \frac{ if_D
\Pi((p+q)^2,P^2,p^2)}{m^2_D-(p-k)^2}
+\int\limits_{s_h^{D}}^{\infty}
ds~\frac{\rho_h^{D}(s,(p+q)^2,P^2,p^2)}{s-(p-k)^2}\,,\label{eq:hadron}
\end{equation}
where $\rho_h^{D}(s,(p+q)^2,P^2,p^2)$ and $s_h^{D}$ are
respectively the spectral function and the threshold mass squared
of the excited and continuum states in $D$ channel,
$\Pi((p+q)^2,P^2,p^2)$ is a two-point correlation function with
the following definition:
\begin{equation}
\Pi((p+q)^2,P^2,p^2)= i\int d^4x\; e^{-i(p+q)x} \langle
D^0(p-k)\pi^0(q)|T\{ \tilde{O}_1(0) j^{(B)}_5(x)\} | 0 \rangle\,.
\label{eq:PI}
\end{equation}

By assuming quark-hadron duality we replace $s_h^{D}$ with the
effective threshold of the perturbative continuum $s_0^{D}$, and
substitute the hadronic spectral density $\rho_{h}^{D}$ in
(\ref{eq:hadron}) with the corresponding QCD one, i.e :
\begin{equation}
\rho_{h}^{D}(s,(p+q)^2,P^2,p^2)\Theta(s-s_h^{D}) = \frac{1}{\pi}
{\rm Im}_s F^{(p-k)}_{QCD}(s,(p+q)^2,P^2,p^2)\Theta(s - s_0^{D})
\, . \label{eq:du1}
\end{equation}
Matching the hadronic relation (\ref{eq:hadron}) onto the QCD
result (\ref{eq:qcd}) yields the expression
\begin{equation}
\frac{i f_D\Pi((p+q)^2,P^2,p^2)}{ m_{D}^2 - (p-k)^2} =
\frac{1}{\pi}\int_{m_c^2}^{s_0^{D}} ds \frac{{\rm Im}_s
F^{(p-k)}_{QCD}(s,(p+q)^2,P^2, p^2)}{s - (p-k)^2} \, .
\label{eq:hadron2}
\end{equation}
which then becomes
\begin{equation}
\Pi((p+q)^2,P^2,p^2)= \frac{-i}{\pi f_D}\int_{m_c^2}^{s_0^{D}}ds~
e^{(m_D^2-s)/M^2}{\rm Im}_{s}
F^{(p-k)}_{QCD}(s,(p+q)^2,P^2,p^2)\,, \label{eq:rPI}
\end{equation}
after the Borel transformation in variable $(p-k)^2$.

Next, for the above expression which is only valid at large
spacellike $P^2$, we have to perform an analytic continuation to
large timelike $P^2$, keeping the variable $(p+q)^2$ fixed. A
natural continuation point is $P^2=m_B^2$. The analytic
continuation of (\ref{eq:rPI} ) yields the result,
\begin{eqnarray}
\Pi((p+q)^2,m_B^2,p^2) & = &
 i\int d^4x\; e^{-i(p+q)x} \langle
D^0(p-k)\pi^0(q)|T\{ \tilde{O}_1(0) j^{(B)}_5(x)\} | 0 \rangle\ \nonumber\\
& = & \frac{-i}{\pi f_D}\int_{m_c^2}^{s_0^{D}}ds~
e^{(m_D^2-s)/M^2}{\rm Im}_{s}
F^{(p-k)}_{QCD}(s,(p+q)^2,m_B^2,p^2)\,. \label{eq:newPI}
\end{eqnarray}
Then we employ the analytical property of the amplitude
$\Pi((p+q)^2,m_B^2,p^2)$ in the spacelike variable $(p+q)^2$.
Inserting in the right hand side of (\ref{eq:newPI}) a complete
set of hadronic states with the $\bar{B}^0$ meson quantum numbers,
we have the following dispersion relation:
\begin{equation}
\Pi((p+q)^2,m_B^2,p^2)= \frac{f_Bm_B^2 \langle D^0(p)\pi^0(q) |O
|\bar{B}^0(p+q) \rangle }{m_B^2-(p+q)^2} +\int_{s_h^{B}}^{\infty}
ds'\frac{\rho_h^{(B)}(s',m_B^2,p^2)}{s'-(p+q)^2}\,,
\label{eq:hdr2}
\end{equation}
where the $B$ meson decay constant is defined as
\begin{equation}
\langle \bar{B}^0|\bar{b}i\gamma_5 d|0\rangle=m_B f_B
\label{eq:fB}.
\end{equation}
At this point, we would like to emphasize that the unphysical
momentum $k$ disappears from the ground state contribution due to
the simultaneous conditions $P^2=(p+q-k)^2=m_B^2$ and
$(p+q)^2=m_B^2$, so that the physical $\bar{B}^0\rightarrow
D^0\pi^0$ matrix element of operator $\tilde{O}_1$ is recovered.

Then (\ref{eq:newPI}) can be changed to a form of double
dispersion relation as
\begin{eqnarray}
\Pi((p+q)^2,m_B^2,p^2)&
=-&\frac{i}{\pi^2f_D}\int_{m_c^2}^{s_0^D}\!ds
\,e^{(m_D^2-s)/M^2}\nonumber\\
&\times& \int_{m_b^2}^{R(s,m_B^2,p^2)}
\frac{ds'}{s'-(p+q)^2}\;\mbox{Im}_{s'} \mbox{Im}_s
F^{(p-k)}_{QCD}(s,s',m_B^2,p^2). \label{eq:Pi2}
\end{eqnarray}
The upper limit R of integration in $s'$ rests generally on $s,\,
m_B^2$ and $p^2$. At present, we make use of quark-hadron duality
once more and approximate the integral in (\ref{eq:hdr2}) by the
$s'\geq s_0^B$ part of the dispersion integral (\ref{eq:Pi2}),
where $s_0^B$ is the effective threshold in the $B$ channel. After
the Borel transformation with respect to the variable $(p+q)^2$ is
made, LCSR for the $\bar{B}^0 \to D^0 \pi^0$ matrix element of the
operator $\tilde{O}_1$ is of the following form:
\begin{eqnarray}
\langle D^0(p-k)\pi^0(q)& |\tilde{O}_1 |& \bar{B}^0(p+q)\rangle =
\frac{-i}{\pi^2f_D f_B m_B^2} \int_{m_c^2}^{s_0^D}\!ds~
e^{(m_D^2-s)/M^2} \nonumber
\\
&\times & \int_{m_b^2}^{\bar{R}(s,m_B^2,p^2,s_0^B)}
\!ds'\;e^{(m_B^2-s')/M'^2}{\rm Im}_{s'}\, {\rm Im}_s
F^{(p-k)}_{QCD}(s,s',m_B^2,p^2)\,, \label{eq:final}
\end{eqnarray}
where $\bar{R}$ is the upper limit of the integration in $s'$
after the use of duality ansatz.

In order to obtain a LCSR estimate of the desired soft
nonfactorizable contribution $\langle
D^0\pi^0|\tilde{O}_1|\bar{B}^0\rangle_S$, which is due to soft
gluon emission off the emitted heavy-light quark pair and
subsequent absorption into the $B\pi$ system, we have to know the
explicit expression of $F^{(p-k)}_{QCD}$. In the derivation of
$F^{(p-k)}_{QCD}$, we employ the light-cone expansion form with
higher-twist terms included for a massive quark propagator, which,
in the fixed-point gauge and only considering correction of
operators with one gluon field, reads \cite{Propagator},
\begin{eqnarray}
S^{ij}(x_1,x_2|m) &\equiv&  -i\langle 0 | T \{q^i(x_1)\,
\bar{q}^j(x_2)\}| 0 \rangle \nonumber
\\
& =& \int\frac{d^4k}{(2\pi)^4}e^{-ik(x_1-x_2)}\Bigg\{
\frac{\not\!k +m}{k^2-m^2} \delta^{ij} -\int\limits_0^1 dv\,  g_s
\, G^{\mu\nu}_a(vx_1+(1-v)x_2) \left (\frac{\lambda^a}{2} \right
)^{ij} \nonumber
\\
& \times &\Big[ \frac12 \frac {\not\!k
+m}{(k^2-m^2)^2}\sigma_{\mu\nu} - \frac1{k^2-m^2}v(x_1-x_2)_\mu
\gamma_\nu \Big]\Bigg\}\,. \label{eq:prop}
\end{eqnarray}
where $G^{\mu\nu}_a$ is the gluon-field strength, and $g_s$ the
strong coupling constant. This means that only the higher-twist
components of light-cone distribution amplitudes for the relevant
pion, which are corresponding to quark-antiquark-gluon nonlocal
operators, would be involved in the final LCSR result.

After some lengthy calculation we obtain the twist-3
contributions:
\begin{eqnarray}
F^{(p-k)}_{tw 3} &=& - \frac{m_b f_{3 \pi}}{4\sqrt{2} \pi^2}
\int_0^1 dv \int {\cal D}\alpha_i \frac{\phi_{3\pi}(\alpha_i,
\mu)} {m_b^2 - (p + q (1-\alpha_1))^2} \, \int_0^1 dx \frac{
x(1-x)}{(1-x) m_c^2
-\bar{Q}^2 x (1-x)} \nonumber \\
& & \times \, q \cdot (p-k) \Big [ (2-v) q \cdot k + 2 (1-v) q
\cdot (p-k) \Big ]\nonumber \\
& &-\frac{m_b f_{3 \pi}}{4\sqrt{2} \pi^2} \int_0^1 dv \int {\cal
D}\alpha_i \frac{\phi_{3\pi}(\alpha_i, \mu)} {m_b^2 - (p + q
(1-\alpha_1))^2} \, \int_0^1 dx \frac{ x(1-x)(2x-1)}{ m_c^2 -
\bar{Q}^2 x (1-x)
} \nonumber \\
& & \times \, q \cdot (p-k) \Big [ (2-3v) q \cdot k + 2 (1-v) q
\cdot (p-k) \Big ]\,. \label{eq:twist3}
\end{eqnarray}
where $\phi_{3\pi}$ is a twist-3 distribution amplitude of pion.
The definition of the twist-3 distribution amplitude as well as of
the twist-4 ones $\phi_{\perp}$, $\phi_{||}$,
$\tilde{\phi}_{\perp}$ and $\tilde{\phi}_{||}$, which will be
encountered in the calculation, is given below through the
relevant matrix elements:
\begin{eqnarray}
-\sqrt{2}\langle 0 |\bar{d}(0) \sigma_{\mu \nu} \gamma_5 G_{\alpha
\beta}(v y) d(x) | \pi^{0}(q) \rangle &=&
i f_{3 \pi} \left [ (q_{\alpha}q_{\mu}g_{\beta \nu} - q_{\beta}q_{\mu}g_{\alpha \nu}) \right .  \nonumber \\
 &&\hspace{-2cm}\left .-  (q_{\alpha}q_{\nu}g_{\beta \mu} -
q_{\beta}q_{\nu}g_{\alpha \mu}) \right ] \int {\cal D}\alpha_i
\phi_{3 \pi}(\alpha_i, \mu) e^{-i q(x \alpha_1 + y v \alpha_3)} ,
\label{eq:tw3a}
\end{eqnarray}
\begin{eqnarray}
& & -\sqrt{2}\langle 0 |\bar{d}(0) i\gamma_{\mu} \tilde{G}_{\alpha
\beta}(v y) d(x) | \pi^{0}(q) \rangle = q_{\mu} \frac{q_{\alpha}
x_{\beta} - q_{\beta}x_{\alpha}}{q x} f_{\pi} \int {\cal
D}\alpha_i \tilde{\phi}_{||}(\alpha_i,\mu)
e^{-i q(x \alpha_1 + y v \alpha_3)} \nonumber \\
& & \hspace{5cm}+ (g_{\mu \alpha}^{\perp}q_{\beta} - g_{\mu
\beta}^{\perp}q_{\alpha})f_{\pi} \int {\cal D}\alpha_i
\tilde{\phi}_{\perp}(\alpha_i,\mu) e^{-i q(x \alpha_1 + y v
\alpha_3)} , \label{eq:tw4a}
\end{eqnarray}
\begin{eqnarray}
& & -\sqrt{2}\langle 0 |\bar{d}(0) \gamma_{\mu} \gamma_5
{G}_{\alpha \beta}(v y) d(x) |\pi^{0}(q) \rangle = q_{\mu}
\frac{q_{\alpha} x_{\beta} - q_{\beta}x_{\alpha}}{q x} f_{\pi}
\int {\cal D}\alpha_i {\phi}_{||}(\alpha_i,\mu)
e^{-i q(x \alpha_1 + y v \alpha_3)} \nonumber \\
& & \hspace{5cm}+ (g_{\mu \alpha}^{\perp}q_{\beta} - g_{\mu
\beta}^{\perp}q_{\alpha}) f_{\pi}\int {\cal D}\alpha_i
{\phi}_{\perp}(\alpha_i,\mu) e^{-i q(x \alpha_1 + y v \alpha_3)}
\, , \label{eq:tw4b}
\end{eqnarray}
where $f_{3\pi}$ is a nonperturbative quantity defined by the
matrix element $\langle 0 |\bar{u}\sigma_{\mu \nu} \gamma_5
G_{\alpha \beta}d|\pi \rangle $, $\tilde{G}_{\alpha \beta} =
\frac{1}{2} \epsilon_{\alpha \beta \rho \sigma} G^{\rho \sigma}$,
$G^{\rho \sigma} = g_s \, \lambda^{a}/2 \, G^{\rho \sigma}_a$,
${\cal D} \alpha_i = d\alpha_1 d\alpha_2 d\alpha_3 \delta( 1 -
\alpha_1 - \alpha_2 - \alpha_3)$, and $g_{\alpha \beta}^{\perp} =
g_{\alpha \beta} - (x_{\alpha} q_{\beta} + x_{\beta}
q_{\alpha})/qx$. The asymptotic forms of all these distribution
amplitudes are given as \cite{WF}
\begin{eqnarray}
\phi_{3\pi}(\alpha_i,\mu)& =& 360 \alpha_1 \alpha_2 \alpha_3^2\, ,
\label{eq:amtw3} \\
\phi_{\perp}(\alpha_i,\mu)& = &10
\delta^2(\mu)(\alpha_1-\alpha_2)\alpha_3^2\, ,
\label{eq:amtw4a} \\
\phi_{||}(\alpha_i,\mu)& =& 120 \delta^2(\mu) \epsilon (\mu)
(\alpha_1-\alpha_2) \alpha_1 \alpha_2 \alpha_3  \, ,
\label{eq:amtw4b} \\
\tilde{\phi}_{\perp}(\alpha_i,\mu)& =& 10 \delta^2(\mu) \alpha_3^2
( 1 - \alpha_3)\, ,
\label{eq:amtw4ta} \\
\tilde{\phi}_{||}(\alpha_i,\mu) &=& -40 \delta^2(\mu) \alpha_1
\alpha_2 \alpha_3\, .
 \label{eq:amtw4tb}
\end{eqnarray}
with $\delta^2(\mu)$ and $\epsilon (\mu)$ being two
nonperturbative parameters.

By changing the order of the integral variables,
(\ref{eq:twist3})\,\,is converted into the following form:
\begin{eqnarray}
F^{(p-k)}_{tw 3} &=&  \frac{m_b f_{3 \pi}}{16\sqrt{2} \pi^2}
\int_{m_c^2}^{\infty} \frac{ds}{s - (p-k)^2}
\int_{\frac{m_c^2}{s}}^{1} dy \int_{x(s,y,P^2)}^1 \frac{du}{m_b^2
- (p+q u)^2}
\nonumber \\
& &\hspace{-2cm} \times \int_{x(s,y,P^2)}^u \frac{dv}{v^2}
\phi_{3\pi}(1 -u, u-v, v) \Big [ s - \frac{m_c^2}{y} + \left (
(p+q)^2 - p^2 \right ) ( 2 v - x(s,y,P^2) ) \Big  ]
\nonumber \\
&+ & \frac{m_b f_{3 \pi}}{16\sqrt{2} \pi^2} \int_{m_c^2}^{\infty}
\frac{ds}{s - (p-k)^2} \int_{\frac{m_c^2}{s}}^{1} dy(2y-1)
\int_{x(s,y,P^2)}^1 \frac{du}{m_b^2 - (p+q u)^2}
\nonumber \\
& &\hspace{-2cm} \times \int_{x(s,y,P^2)}^u \frac{dv}{v^2}
\phi_{3\pi}(1 -u, u-v, v)  \Big [  \frac{m_c^2}{y}-s + \left (
(p+q)^2 - p^2 \right ) ( 2 v - 3x(s,y,P^2) ) \Big  ]\, ,
\label{eq:tw3}
\end{eqnarray}
where $x=(s-\frac{m_c^2}{y})/(s-P^2)$.

The derivation of twist-4 contributions turn out to be even more
tedious. The result is given as follows
\begin{eqnarray}
F^{(p-k)}_{tw 4} &=& \frac{m_b^2 f_{\pi}}{8\sqrt{2} \pi^2}
\int_{m_c^2}^{\infty} \frac{ds}{s - (p-k)^2}
\int_{\frac{m_c^2}{s}}^{1} dy \int_{x(s,y,P^2)}^1 \frac{du}{m_b^2
- (p+q u)^2}
\nonumber \\
& & \hspace*{-1cm}\times \int_{x(s,y,P^2)}^u \frac{dv}{v}
\tilde{\phi}_{\perp}(1 -u, u-v, v) \left [ 3 - \frac{2}{v}
x(s,y,P^2) \right ]
\nonumber \\
& & + \frac{m_b^2 f_{\pi}}{8\sqrt{2} \pi^2} \int_{m_c^2}^{\infty}
\frac{ds}{s - (p-k)^2} \int_{\frac{m_c^2}{s}}^{1} dy(2y-1)
\int_{x(s,y,P^2)}^1 \frac{du}{m_b^2 - (p+q u)^2}
\nonumber \\
& & \hspace*{-1cm}\times \int_{x(s,y,P^2)}^u \frac{dv}{v}
{\phi}_{\perp}(1 -u, u-v, v) \left [ 3 - \frac{4}{v} x(s,y,P^2)
\right ]
\nonumber \\
& &  \hspace*{-2cm} + \frac{m_b^2 f_{\pi}}{8\sqrt{2} \pi^2}
\int_{m_c^2}^{\infty} \frac{ds}{s - (p-k)^2}
\int_{\frac{m_c^2}{s}}^{1} dy \int_{x(s,y,P^2)}^1 \frac{du}{\left
[ (m_b^2 - (p+q u)^2 \right ]^2}
\nonumber \\
& & \hspace*{-1cm}\times \int_{x(s,y,P^2)}^u \frac{dv}{v^2}
\tilde{\Phi}_{1}(1 -u, v) \left [ s - \frac{m_c^2}{y} + ((p+q)^2 -
p^2) ( -v + x(s,y,P^2) )\right ]
\nonumber \\
& &  \hspace*{-2cm} - \frac{m_b^2 f_{\pi}}{8 \sqrt{2}\pi^2}
\int_{m_c^2}^{\infty} \frac{ds}{s - (p-k)^2}
\int_{\frac{m_c^2}{s}}^{1} dy(2y-1) \int_{x(s,y,P^2)}^1
\frac{du}{\left [ (m_b^2 - (p+q u)^2 \right ]^2}
\nonumber \\
& & \hspace*{-1cm}\times \int_{x(s,y,P^2)}^u \frac{dv}{v^2}
\Phi_{1}(1 -u, v) \left [ 2(s - \frac{m_c^2}{y}) + (P^2 - s)v
\right ]
\nonumber \\
& & \hspace*{-2cm} - \frac{m_b^2 f_{\pi}}{8 \sqrt{2}\pi^2}
\int_{m_c^2}^{\infty} \frac{ds}{s - (p-k)^2}
\int_{\frac{m_c^2}{s}}^{1} dy \int_{x(s,y,P^2)}^1 \frac{du}{\left
[m_b^2 - (p+q u)^2 \right ]^2}
\nonumber \\
& & \hspace*{-1cm}\times \frac{\tilde{\Phi}_{2}(u)}{u^2} \left [ s
- \frac{m_c^2}{y} + ((p+q)^2 - p^2) (-u + x(s,y,P^2) \right ]
\nonumber \\
& & \hspace*{-2cm} - \frac{m_b^2 f_{\pi}}{8 \sqrt{2}\pi^2}
\int_{m_c^2}^{\infty} \frac{ds}{\left [ s - (p-k)^2 \right ]^2}
\int_{\frac{m_c^2}{s}}^{1} dy \frac{P^2}{P^2 - \frac{m_c^2}{y}}
\int_{x(s,y,P^2)}^1 \frac{du}{m_b^2 -
(p+qu)^2} \nonumber \\
 & & \hspace*{-1cm} \times
\frac{\tilde{\Phi}_{2}(u)}{u^2} \left [
\frac{x(s,y,P^2)}{-P^2}\left (2\, q \cdot (p-k) \right )^2 \left
(1 -\frac{x(s,y,P^2)}{u}\frac{q\cdot k}{q \cdot p}\right ) \right
]\nonumber \\
& & \hspace*{-2cm} + \frac{m_b^2 f_{\pi}}{8 \sqrt{2}\pi^2}
\int_{m_c^2}^{\infty} \frac{ds}{s - (p-k)^2}
\int_{\frac{m_c^2}{s}}^{1} dy(2y-1) \int_{x(s,y,P^2)}^1
\frac{du}{\left [m_b^2 - (p+q u)^2 \right ]^2}
\nonumber \\
& & \hspace*{-1cm}\times \frac{\Phi_{2}(u)}{u^2} \left [2( s -
\frac{m_c^2}{y}) + (P^2 - s) u  \right ]
\nonumber \\
& & \hspace*{-2cm} - \frac{m_b^2 f_{\pi}}{8 \sqrt{2}\pi^2}
\int_{m_c^2}^{\infty} \frac{ds}{\left [ s - (p-k)^2 \right ]^2}
\int_{\frac{m_c^2}{s}}^{1} dy(2y-1)
\frac{P^2}{P^2-\frac{m_c^2}{y}} \int_{x(s,y,P^2)}^1
\frac{du}{m_b^2 - (p+q u)^2}
\nonumber \\
& & \hspace*{-1cm} \times \frac{\Phi_{2}(u)}{u^2} \left [
\frac{x(s,y,P^2)}{-P^2}\left (2\, q \cdot (p-k) \right )^2 (1
-\frac{2x(s,y,P^2)}{u})(1-\frac{q\cdot k}{q \cdot p}) \right ] \,
 \label{eq:tw4}
\end{eqnarray}
with the scalar functions $\Phi_i$'s and $\tilde{\Phi}_i$'s
defined by
\begin{eqnarray}
\Phi_1(u,v) &=& \int_0^u d\omega \left ( \phi_{\perp}(\omega, 1 -
\omega - v,v) + \phi_{||}(\omega, 1 - \omega - v,v) \right )\, ,
\nonumber \\
\hspace*{-0.5cm} \Phi_2(u) &=& \int_0^u d\omega'
\int_0^{1-\omega'} d \omega'' \left ( \phi_{\perp}(\omega'', 1 -
\omega'' -\omega', \omega') + \phi_{||}(\omega'', 1 - \omega''
-\omega', \omega') \right ),\\
 \tilde{\Phi}_1(u,v)
&=&\int_0^ud\omega \left ( \tilde{\phi}_{\perp}(\omega, 1 - \omega
- v,v) + \tilde{\phi}_{||}(\omega, 1 - \omega - v,v) \right )\, ,
\nonumber \\
\hspace*{-0.5cm} \tilde{\Phi}_2(u) &=& \int_0^u d\omega'
\int_0^{1-\omega'} d \omega'' \left (
\tilde{\phi}_{\perp}(\omega'', 1 - \omega'' -\omega', \omega') +
\tilde{\phi}_{||}(\omega'', 1 - \omega'' -\omega', \omega') \right
).
\label{eq:twist4}
\end{eqnarray}
The resulting expression (\ref{eq:tw4}) is much more complicated
than those in the case of $B\longrightarrow\pi\pi, \pi K$
\cite{LCSR1} and $B\longrightarrow J/\psi K $\cite{LCSR2}, because
of the mass asymmetry of the two quarks in $D$ meson. In contrast
to the latter case in which there is no contribution of $\phi_i$'s
due to cancellation in the corresponding twist-4 parts, the
twist-4 piece receive the contributions from not only
$\tilde{\phi}_i$'s but also $\phi_i$'s in the present case.

To proceed, we should change the above expressions (\ref{eq:tw3})
and (\ref{eq:tw4}) into the desired form of double dispersion
relation. For the twist-3 contribution $F^{(p-k)}_{tw3}$, we have
a dispersion expression in $(p-k)^2$:
\begin{equation}
F^{(p-k)}_{tw 3} = \frac{1}{\pi} \int_{m_c^2}^{\infty} \frac{ds}{s
- (p-k)^2} {\rm Im_s}\, F_{tw3}^{(p-k)} (s, (p+q)^2, P^2,p^2)\, ,
\label{eq:tw3im1}
\end{equation}
where
\begin{eqnarray}
{\rm Im_s}\, F_{tw3}^{(p-k)} (s, (p+q)^2, P^2,p^2)
&=& \nonumber \\
& & \hspace*{-5cm}   \frac{m_b f_{3 \pi}}{16 \sqrt{2}\pi}
\int_{\frac{m_c^2}{s}}^{1}dy \int_{x(s,y,P^2)}^1 \frac{du}{m_b^2 -
(p+q u)^2} \int_{x(s,y,P^2)}^u \frac{dv}{v^2} \phi_{3\pi}(1 -u,
u-v, v)
\nonumber \\
& & \hspace*{-5cm} \times \Big [ s - \frac{m_c^2}{y} + \left (
(p+q)^2 - p^2 \right ) ( 2 v - x(s,y,P^2) ) \Big  ] \nonumber \\
& & \hspace*{-5cm} + \frac{m_b f_{3 \pi}}{16\sqrt{2} \pi}
\int_{\frac{m_c^2}{s}}^{1}dy(2y-1) \int_{x(s,y,P^2)}^1
\frac{du}{m_b^2 - (p+q u)^2} \int_{x(s,y,P^2)}^u \frac{dv}{v^2}
\phi_{3\pi}(1 -u, u-v, v)
\nonumber \\
& & \hspace*{-5cm} \times \Big [ -s+ \frac{m_c^2}{y} + \left (
(p+q)^2 - p^2 \right ) ( 2 v - 3x(s,y,P^2) ) \Big  ]\, .
\label{eq:tw3im2}
\end{eqnarray}
Then for ${\rm Im_s}\, F^{(p-k)}_{tw3}$ we make Tailor expansion
in variable $x(s,y,P^2)$. Up to order ${\cal O}(x^3)$ the result
is
\begin{eqnarray}
{\rm Im_s} F_{tw3}^{(p-k)}(s,(p+q)^2,P^2) &=&
 \nonumber \\
 & & \hspace*{-5cm} \frac{m_b f_{3 \pi}}{16\sqrt{2} \pi^2} \int_0^1 \frac{du}{m_b^2 - (p+q u)^2}
\nonumber \\
 & &\hspace*{-5cm} \times \int_{\frac{m_c^2}{s}}^{1} dy
 \Bigg\{ \int_0^u \frac{dv}{v^2} \phi_{3\pi}(1 -u, u-v, v)
 \left [ s -  \frac{m_c^2}{y} + 2 v ((p+q)^2 - p^2) \right ] \nonumber \\
 & & \hspace*{-4cm}
 - \left [ \int_0^u \frac{dv}{v^2} \phi_{3\pi}(1 -u, u-v, v) \left ( (p+q)^2 - p^2 \right )
\right . \nonumber \\
 & & \hspace*{-4cm}  \left .  + \left ( s -  \frac{m_c^2}{y} \right ) \left ( \frac{1}{v^2} \phi_{3\pi}(1 -u, u-v, v) \right )_{v=0}
 \right ] x(s,y,P^2) \nonumber \\
 & &  \hspace*{-5cm} -
  \left ( s -  \frac{m_c^2}{y} \right )
 \left [ \frac{\partial}{\partial v} \left ( \frac{1}{v^2} \phi_{3\pi}(1 -u, u-v, v) \right ) \right ]_{v=0}
 \frac{x^2(s,y,P^2)}{2} \Bigg\}  \nonumber \\
& &\hspace*{-5cm} +\frac{m_b f_{3 \pi}}{16\sqrt{2} \pi^2} \int_0^1 \frac{du}{m_b^2 - (p+q u)^2} \nonumber \\
 & &\hspace*{-5cm} \times \int_{\frac{m_c^2}{s}}^{1} dy(2y-1)
 \Bigg\{ \int_0^u \frac{dv}{v^2} \phi_{3\pi}(1 -u, u-v, v)
 \left [ -s+\frac{m_c^2}{y} + 2 v ((p+q)^2 - p^2) \right ] \nonumber \\
 & & \hspace*{-4cm}
 - \left [ 3 \int_0^u \frac{dv}{v^2} \phi_{3\pi}(1 -u, u-v, v) \left ( (p+q)^2 - p^2 \right )
\right . \nonumber \\
 & & \hspace*{-4cm}  \left .  + \left ( -s+ \frac{m_c^2}{y} \right ) \left ( \frac{1}{v^2} \phi_{3\pi}(1 -u, u-v, v) \right )_{v=0}
 \right ] x(s,y,P^2) \nonumber \\
 & &  \hspace*{-5cm} +\left [  \left ( \frac{4}{v^2} \phi_{3\pi}(1 -u, u-v, v) \left ( (p+q)^2 - p^2 \right ) \right)\
\right . \nonumber \\
& & \hspace*{-5cm} - \left . \left ( -s+  \frac{m_c^2}{y} \right )
  \frac{\partial}{\partial v} \left ( \frac{1}{v^2} \phi_{3\pi}(1 -u, u-v, v) \right ) \right ]_{v=0}
 \frac{x^2(s,y,P^2)}{2} \Bigg\}
  +  {\cal O}(x^3) \, .
 \label{eq:tw3exp}
 \end{eqnarray}
With the substitution $u=(m_b^2-p^2)/(s'-p^2)$, the integral in
$u$ in the above equation can get back to its dispersion form
\begin{equation}
\int_{0}^{1}du\frac{F(u)}{m_b^2-(p+qu)^2}=\int_{m_b^2}^{\infty}
\frac{ds'}{s'-(p+q)^2}\frac{F(u(s'))}{s'-p^2}\, .
\end{equation}
At last the desired double dispersion form is achieved.

The twist-4 contribution $F^{(p-k)}_{tw4}$ in (\ref{eq:tw4}) can
be treated similarly. The derivation is omitted to save some
space. We note that compared with the resulting twist-3
contribution, the twist-4 part has some additional terms
containing denominators of the form
\begin{equation}
\frac{1}{\left [ s - (p-k)^2 \right ]^2} \qquad {\rm or}\qquad
\frac{1}{\left [ m_b^2 - (p+u q)^2 \right ]^2} \, .
\end{equation}
As argued in \cite{LCSR2}, however, these terms containing higher
power of such denominators are numerically suppressed and can be
neglected safely. Therefore in the ensuing discussion we will not
take them into account.

Putting everything together, we obtain the final LCSR result for
the soft contribution to the matrix element $\langle D^0(p)
\pi^0(q) | \tilde{O}_1(0) |\bar{B}^0(p+q) \rangle$:
\begin{eqnarray}
\langle D^0(p) \pi^0(q) | \tilde{O}_1(0) |\bar{B}^0(p+q) \rangle_S &=&\frac{-i m_b}{8\sqrt{2}\pi^2 f_D f_B m_B^2} \nonumber\\
& &\hspace{-5cm}\times \int_{m_c^2}^{s_0^{D}} ds
e^{(m_D^2-s)/M^2}\int_{u_0^B}^1 \frac{du}{u} e^{(m_B^2-(m_b^2 -
m_{D}^2(1-u))/u)/M'^2} \nonumber\\
& & \hspace*{-6cm}\times \int_{\frac{m_c^2}{s}}^{1} dy \Bigg \{
\frac{f_{3\pi}}{2} \Bigg [ \int_0^u \frac{dv}{v^2} \phi_{3\pi}(1
-u, u-v, v)
 \left ( \frac{m_b^2 - m_{D}^2}{u} ( 2 v -  x(s,y,m_B^2) ) + s -  \frac{m_c^2}{y} \right )  \nonumber \\
 & & \hspace*{-5cm}  - \left ( s -  \frac{m_c^2}{y} \right ) \left ( \frac{1}{v^2} \phi_{3\pi}(1 -u, u-v, v) \right )_{v=0}
  x(s,y,m_B^2) \nonumber \\
 & &  \hspace*{-5cm} -
  \left ( s -  \frac{m_c^2}{y} \right )
 \left [ \frac{\partial}{\partial v} \left ( \frac{1}{v^2} \phi_{3\pi}(1 -u, u-v, v) \right ) \right ]_{v=0}
 \frac{x^2(s,y,m_B^2)}{2} \Bigg ]
\nonumber \\
& & \hspace{-6cm}+ m_b f_\pi \left [\int_0^u \frac{dv}{v^2}
\tilde{\phi}_{\perp}(1 -u, u-v, v) \left( 3 -   \frac{2}{v}
x(s,y,m_B^2)\right)\nonumber \right . \\
& & \hspace{-5cm}\left .
+\left(\frac{3}{v^2}\tilde{\phi}_{\perp}(1 -u, u-v, v)
-\left(\frac{1}{v}\frac{\partial}{\partial v }
\tilde{\phi}_{\perp}(1 -u, u-v, v)
\right)\right)_{v=0}\frac{x^2(s,y,m_B^2)}{2}\right ] \Bigg
\}\nonumber \\
& &\hspace{-6cm} + \frac{-i m_b}{8\sqrt{2}\pi^2 f_D f_B m_B^2}
\int_{m_c^2}^{s_0^{D}} ds e^{(m_D^2-s)/M^2}\int_{u_0^B}^1
\frac{du}{u} e^{(m_B^2-(m_b^2 - m_{D}^2(1-u))/u)/M'^2}
\int_{\frac{m_c^2}{s}}^{1} dy(2y-1)
 \nonumber \\
& & \hspace*{-5.5cm} \times \Bigg \{ \frac{f_{3\pi}}{2} \Bigg [
 \int_0^u \frac{dv}{v^2} \phi_{3\pi}(1 -u, u-v, v)
 \left ( \frac{m_b^2 - m_{D}^2}{u} ( 2 v -  3x(s,y,m_B^2) ) - s + \frac{m_c^2}{y} \right )  \nonumber \\
 & & \hspace*{-5cm}  - \left ( -s +  \frac{m_c^2}{y} \right ) \left ( \frac{1}{v^2} \phi_{3\pi}(1 -u, u-v, v) \right )_{v=0}
  x(s,y,m_B^2) \nonumber \\
 & &  \hspace*{-5cm} + \left [ 4\frac{m_b^2-m_D^2}{u}\frac{1}{v^2}\phi_{3\pi}(1 -u, u-v, v) \right .
\nonumber \\
& &  \hspace*{-5cm} \left . -  \left (- s +  \frac{m_c^2}{y}
\right ) \frac{\partial}{\partial v} \left ( \frac{1}{v^2}
\phi_{3\pi}(1 -u, u-v, v) \right ) \right ]_{v=0}
 \frac{x^2(s,y,m_B^2)}{2} \Bigg ]
\nonumber \\
& & \hspace{-6cm}+ m_b f_\pi \left [\int_0^u \frac{dv}{v^2}
\tilde{\phi}_{\perp}(1 -u, u-v, v) \left( 3 -   \frac{4}{v}
x(s,y,m_B^2)\right)\nonumber \right . \\
& & \hspace{-6cm}\left . +\left (3\left
(\frac{1}{v^2}\tilde{\phi}_{\perp}(1 -u, u-v, v)\right
 )+\left (\frac{1}{v}\frac{\partial}{\partial v
} \tilde{\phi}_{\perp}(1 -u, u-v, v)\right ) \right
)_{v=0}\frac{x^2(s,y,m_B^2)}{2}\right ] \Bigg \} \Bigg \} \, ,
\label{eq:finalresult}
\end{eqnarray}
where $u_0^B = (m_b^2 - m_{D}^2)/(s_0^B - m_{D}^2)$.

Let's proceed to numerical discussion. The $D$ channel parameters
are taken as \cite{Belyaev} $m_D=1.87$ GeV, $m_c=1.3\pm 0.1$ GeV,
$f_D=170\pm 10$ MeV $, s_0^D=6\pm 1$ GeV$^2$ and $M^2=1.5\pm 0.5$
GeV$^2$. The parameters in $B$ channel are chosen as
\cite{Input1}: $m_B=5.28$ GeV, $m_b=4.7\pm 0.1$ GeV, $f_B=180\pm
30$ GeV, $s_0^B=35\pm 2$ GeV$^2$ and $M'^2=10\pm 2$ GeV$^2$. For
the nonperturbative quantities entering the relevant light-cone
distribution amplitudes, we use \cite{Belyaev} $f_{3 \pi} =
0.0026$ GeV$^2$, $\delta^2(\mu_b) = 0.17$ GeV$^2$ and $\epsilon
(\mu_b)=0.36$, whth $\mu_b = \sqrt{m_B^2 - m_b^2} \sim m_b/2 \sim
2.4$ GeV. With these inputs, the contributions of twist-3 and -4
fall into the following ranges respectively:
\begin{equation}
i\langle D^0 \pi^0|\tilde{O_1}|\bar{B}^0\rangle_S ^{(tw3)}=(0.024-
0.053) \textsl{GeV\,}^3,
\end{equation}
and
\begin{equation}
i\langle D^0 \pi^0|\tilde{O_1}|\bar{B}^0\rangle_S ^{(tw4)}=(0.009-
0.017) \textsl{GeV\,}^3,
\end{equation}
the total contribution reads
\begin{equation}
i\langle D^0 \pi^0|\tilde{O_1}|\bar{B}^0\rangle_S =(0.033- 0.070)
\textsl{GeV\,}^3. \label{eq:finalnum}
\end{equation}
These sum rule results show a good stability against the
variations of both the Borel parameters in the given ranges.

\section{Discussion and Conclusion}

Having at hand the LCSR result (\ref{eq:finalnum}) for the matrix
element $\langle D^0 \pi^0|\tilde{O_1}|\bar{B}^0\rangle_S $, we
can discuss the numerical influence of the power-suppressed soft
effect on $\bar{B}^0\to D^0\pi^0$.

Taking $C_1(\mu_b)=-0.257, C_2(\mu_b)=1.117$, $|V_{cb}|=0.043$ and
$|V_{ud}|=0.974$, the magnitude of ${\cal A_S}$ given by
(\ref{eq:softnonfctorizable}) is estimated at
\begin{equation}
|{\cal A}_{S}(\bar{B}^0\longrightarrow
D^0\pi^0)|=(2.48-5.27)\times 10^{-8} \textsl{GeV}.
\end{equation}
It is in order that we make a numerical comparison between ${\cal
A_S}(\bar{B}^0\longrightarrow D^0\pi^0)$  and the naive
factorization piece of the decay amplitude ${\cal
A}_F(\bar{B}^0\longrightarrow D^0\pi^0)$ given by
(\ref{eq:factorizable}). With the LCSR result
$F_0^{B\pi}(m_D^2)=0.30$ \cite{Input2}, we have
\begin{eqnarray}
R_1&=&{\cal A}_{S}(\bar{B}^0\longrightarrow D^0\pi^0)/{\cal
A}_F(\bar{B}^0\longrightarrow D^0\pi^0).\nonumber \\
&=&0.54-1.15.
\end{eqnarray}
This result shows that the resulting soft effect is comparable
numerically with the corresponding factorizable one. A similar
ratio was estimated for the $B\to J/\psi K $ decay in
Ref.\cite{LCSR2}, with the value $0.30-0.70$. It seems that power
suppressed soft effects are even more important in $(\bar{B}^0\to
D^0\pi^0)$ than in $B\to J/\psi K $, as expected.

It is also interesting to compare numerically ${\cal
A}_{S}(\bar{B}^0\longrightarrow D^0\pi^0)$ with the factorizable
contribution of the $O_2$ operator, ${\cal
A}_F^{(O_2)}(\bar{B}^0\longrightarrow D^0\pi^0)=-\frac{i}{6} C_2
G_F V_{cb}V^*_{ud}m^2_B f_D F_0^{B\pi}(m^2_D)$. To this end, we
estimate the ratio
\begin{equation}
R_2={\cal A}_{S}(\bar{B}^0\longrightarrow D^0\pi^0)/{\cal
A}_F^{(O_2)}(\bar{B}^0\longrightarrow D^0\pi^0). \nonumber
\end{equation}
The result is $R_2=0.17-0.35$. Explicitly, our LCSR calculations
favor $a_2^{NL}>0$, indicating that the correction to $a_2$, which
is relevant to the factorizable and power-suppressed soft part, is
positive. This forms a striking contrast to the case of
Ref.\cite{Halperin} where $R_2$ is found to be -0.7, a large
negative number, so that sign of $a_2^{NL}$ is predicted to be
negative, i.e, $a_2$ would receive a negative number correction
from such non-leading effects.

Naive factorization does not apply for the color-suppressed
$\bar{B}^0\longrightarrow D^0\pi^0$ decay and as a consequence,
the power-suppressed soft exchange correction is expected to be
important and is worth discussing carefully, in spite of its
non-leading character. We discuss such effect in the generalized
QCD LCSR. The numerical result is in agreement with one's
expectations. The size of the resulting contribution to the decay
amplitude is found to be comparable with the corresponding
factorizable one, about $(50-110)\%$ of the latter, and the
parameter $a_2$ would receive a positive number correction,
analogously to the case of $B\to J/\psi K$. These observations
would be of important phenomenological interests. Of course, at
this stage we are not able to go a step further to give a complete
estimate of the $\bar{B}^0\longrightarrow D^0\pi^0$ decay
amplitude, due to the unknown leading nonfactorizable soft
contributions. More theoretical or phenomenological efforts in
this direction are necessary to better understand the
color-suppressed charmed decays of $B$ mesons.


\begin{thebibliography}{99}


\bibitem{nf} M. Bauer, B. Stech, M. Wirbel,
Z. Phys. C {\bf 29}, 637 (1985); {\sl ibid.} {\bf 34}, 103 (1987).
\bibitem{gf} M. Neubert, V. Rieckert, B. Stech, Q.P. Xu, Heavy Flavors, ed. by
A.J. Buras, M. Lindner (World Scientific, Singapore,1992), pp.286;
M. Neubert, B. Stech, Adv. Ser. Direct. High Energy Phys. {\bf
15}, 294 (1998); A. Ali and C. Greub, Phys. Rev. D {\bf 57}, 1996
(1998); A. Ali, G. Kramer and C. D. Lu, \textit{ibid}. {\bf 58},
094009 (1998); H.Y. Cheng and B. Tseng, \textit{ibid}. {\bf 58},
094005 (1998).
\bibitem{qcdfa} M. Beneke, G. Buchalla, M. Neubert, and C.T. Sachrajda,
Phys. Rev. Lett. {\bf 83}, 1914 (1999); Nucl. Phys. B {\bf 591},
313 (2000).
\bibitem{bjorken} J.D. Bjorken, Nucl. Phys. (Proc.Suppl) B {\bf 11}, 325 (1989).
\bibitem{Buras} A.J. Buras, J.M. Gerard, R. R\"{u}ckle, Nucl. Phys. B {\bf 268}, 16 (1986).
\bibitem{Donoghue} J.F. Donoghue, E. Golowich, A.A. Petrov and
J.M. Soares, Phys. Rev. Lett. {\bf 77}, 2178 (1996); B. Blok amd
I. Halpern, Phys. Lett. B {\bf 385}, 324 (1996).
\bibitem{CLEO} CLEO Colla., T.E. Coan {\it et al.}, Phys. Rev. Lett. {\bf 88}, 062001 (2002).
\bibitem{Belle} Belle Colla., K. Abe {\it et al.}, Phys. Rev. Lett. {\bf 88}, 052002 (2002).
\bibitem{Keum}  Y.Y. Keum, T. Kurimoto, H.N. Li, C.D. Lu and A.I. Sada, Phys. Rev. D {\bf 69}, 094018 (2004).
\bibitem{scet}S. Mantry, D. Pirjol and I. W. Stewart, Phys. Rev. D {\bf 68}, 114009 (2003).
\bibitem{neubert}  M. Neubert, A.A. Petrov, Phys. Lett. B {\bf 519}, 50 (2001).
\bibitem{xing} Z.Z. Xing, hep-ph/0107257.
\bibitem{cheng} H.Y. Cheng, Phys. Rev. D {\bf 65}, 094012 (2002).
\bibitem{LCSR}  I.I. Balitsky, V.M. Braun and A.V. Kolesnichenko, Nucl. Phys. B {\bf 312}, 509 (1989);
V.M. Braun and I.E. Filyanov, Z. Phys. C {\bf 44}, 157 (1989); V.
L. Chernyak and I.R. Zhitnitsky,  Nucl. Phys. B {\bf 345}, 137
(1990).
\bibitem{Halperin} I. Haperin, Phys. Lett. B {\bf 349}, 548 (1995).
\bibitem{Kh} A. Khodjamirian, Nucl. Phys. B {\bf 558}, 605 (2001).
\bibitem{LCSR1} X.Y. Wu, Z.H. Li, C.J. Cui and T. Huang, Chin. Phys. Lett. {\bf 19}, 1596 (2002); A. Khodjamirian,
T. Mannel and P. Urban, Phys. Rev. D {\bf 67}, 054027 (2003); A.
Khodjamirian, T. Mannel and B. Meli\'{c}, hep-ph/0304179; T.
Huang, L. Li, X.Q. Li, Z.H. Li and X.Y. Wu, hep-ph/0404149.
\bibitem{LCSR2} B. Meli\'{c}, Phys. Rev. D {\bf 68}, 034004 (2003).
\bibitem{Hamiltonian}G. Buchalla, A.J. Buras and M.E. Lautenbacher, Rev. Mod. Phys. {\bf 68}, 1125 (1996).
\bibitem{Propagator} J. Bijnens and A. Khodjamirian, Eur. J. C {\bf 26}, 67 (2002)
\bibitem{WF} V.M. Braun and I.F. Filyanov, Z. Phys. C {\bf 48}, 239
(1990); P. Ball, JHEP {\bf 01}, 010 (1999).
\bibitem{Belyaev} V.M. Belyaev, V.M. Braun, A. Khodjamirian and R.
R\"{u}ckl, Phys. Rev. D {\bf 51}, 6177 (1995).
\bibitem{Input1}A. Khodjamirian, R. R\"{u}ckl, S. Weinzierl, C.W.
Winhart and O.Yakovlev, Phys.Rev. D {\bf 62}, 114002 (2000).
\bibitem{Input2}A. Khodjamirian, R. Ruckl, C.W. Winhart, Phys. Rev. D{\bf58}
054013 (1998).


\end{thebibliography}
\end{document}